\documentclass[preprint,12pt]{elsarticle}

\usepackage{array}
\usepackage{amssymb,amsmath}
\usepackage{mathtools}
\usepackage{nicefrac}
\usepackage{booktabs}
\usepackage{soul}
\usepackage{color}
\usepackage{algorithm}
\usepackage{algpseudocode}

\journal{Information Fusion}

\renewcommand{\hl}[1]{#1}
\newcommand{\hll}[1]{#1}

\newcommand*{\defeq}{\vcentcolon=}
\newcommand{\bb}[1]{\mathbf{#1}}
\newcommand*{\tr}{\operatorname{tr}} 
\newcommand*\diff{\mathop{}\!\mathrm{d}}

\newcommand*{\Cov}{\operatorname{Cov}}

\begin{document}

\begin{frontmatter}



\title{Analytic solution of the exact Daum–Huang flow equation for particle filters}


\author[kjit]{Olivér Törő}
\author[kjit]{Tamás Bécsi}

\address[kjit]{Department of Control for Transportation and Vehicle Systems, Faculty of Transportation Engineering and Vehicle Engineering, Budapest University of Technology and Economics,
            Műegyetem rkp. 3., Budapest, H-1111, Hungary}

\begin{abstract}
State estimation for nonlinear systems, especially in high dimensions, is a generally intractable problem, despite the ever-increasing computing power. Efficient algorithms usually apply a finite-dimensional model for approximating the probability density of the state vector or treat the estimation problem numerically. In 2007 Daum and Huang introduced a novel particle filter approach that uses a homotopy-induced particle flow for the Bayesian update step. Multiple types of particle flows were derived since with different properties. The exact flow considered in this work is a first-order linear ordinary time-varying inhomogeneous differential equation for the particle motion. An analytic solution in the interval [0,1] is derived for the scalar measurement case, which enables significantly faster computation of the Bayesian update step for particle filters.
\end{abstract}



\begin{keyword}
Particle filter \sep Particle flow \sep State estimation \sep Nonlinear filtering \sep Log-homotopy


\end{keyword}

\end{frontmatter}


\section{Introduction}
\label{sec:intro}
Nonlinear state estimation is, in general, not a tractable problem and keeps attracting focus. In the Bayesian framework, two main approaches can be identified. For arbitrary density functions, general numeric solutions can be used, which may provide optimal performance at a huge computational cost. On the other hand, if some specific distributi on or system dynamic is assumed, a more specific estimator could be used. The most well-known specialized estimator is the Kalman filter (KF) which works in the linear Gaussian regime and gives an optimal performance with low computational requirement \cite{kalman1960}. For nonlinear systems, different variants of the Kalman filter are at hand, including the Extended Kalman filter (EKF) \cite{ljung1979asymptotic}, Unscented Kalman filter (UKF) \cite{julier1997new}, Cubature Kalman filter (CKF) \cite{arasaratnam2009cubature}, and Ensemble Kalman filter (EnKF) \cite{evensen2003ensemble}. For multimodal distributions, Gauss Mixture approaches are available \cite{stordal2011bridging}. One important property of these filters is that the distribution remains Gaussian during the estimation, which means only the mean and the covariance need to be propagated. This property makes these filters finite-dimensional, which is a must-have feature for practical estimators. Finite-dimensional filters exist for non-Gaussian problems also, given the distribution is from the exponential family and some requirements towards the system dynamics are met. A summary of these types of filters can be found in \cite{daum2005nonlinear}.

Regarding numerical approaches, the particle filter can be used for arbitrary distributions and system models. It approximates the probability densities by Monte Carlo sampling hence its alternative name, sequential Monte Carlo estimator. The original particle filter, published in \cite{gordon1993novel} is easy to implement; however, it does not give satisfactory performance. A particle filter to run effectively needs careful design and much attention to monitor the quality of the particle ensemble at runtime \cite{ristic2003beyond}. To name a few aspects, one needs to create an adequate proposal distribution, choose a sampling method, design a resampling strategy, adjust the particle number at runtime, deal with particle depletion, regularization and on top of them comes the curse of dimensionality, which is the exponential growth of the needed computation, or equivalently particle number, with the dimension \cite{daum2003curse}.

To address some particle filter related issues, Daum and Huang proposed a new approach in \cite{daum2007nonlinear}. The insight was that what we are bad at is not the prediction but the update step, which needs to be implemented more efficiently in a progressive manner. This approach involves particles not to be weighted or resampled but moved to the proper location in the state space. \hl{The motion is induced by a homotopy equation based on the logarithmic Bayes' rule.}
Contrary to the particle filters, the Daum–Huang type log-homotopy particle flow filter needs much less maintenance, has higher computational complexity per particle, and uses much fewer particles.

The concept of a progressive update is not without history; previous approaches, however, did not use log-homotopy and, more importantly, are different in nature. \hll{Oudjane and Musso introduced the progressive Bayesian update for regularized particle filters in \cite{oudjane2000progressive}. Their approach is to factorize the likelihood function in a principled way to minimize the cost coefficients of the subupdate steps. As the cost is defined in a way that it measures the discrepancy between the prior and the likelihood function, the proposed method aims to perform the update step progressively such that the particle degeneracy originating from a narrow likelihood function is minimized.} In \cite{hanebeck2003progressive} a progressive update method of the probability density is used, which is achieved by a system of linear first-order ordinary differential equations (ODE). The independent parameter of the differential equations starts from 0 and increases up to one. The squared integral deviation between the true and the approximated density defines the dynamics that govern the evolution of the density function.

In \cite{huber2013gaussian} polynomial nonlinearities and exponential distributions are considered giving rise to a homotopy-based moment calculation via ODEs.
Another approach includes sub-likelihoods for the progressive steps, deterministic samples, and particle weights \cite{hanebeck2016progressive}. \hl{Additional particle flow type estimators can be found in \cite{heng2021gibbs} or \cite[Chapter~9.3]{evensen2022data}.}

This paper considers the exact Daum–Huang (EDH) particle flow, which is a deterministic flow described by first-order differential equations. \hl{The standard approach of the particle flow filter is to numerically integrate the flow equation to obtain the posterior distribution from the prior, which is, of course, a computationally intensive practice and gives approximate results. To this end, the paper proposes an analytical solution based method in the interval [0,1], which corresponds to the Bayesian update step for particle filters. First, the commutative property of the differential equation is verified then the solution and its derivation for the scalar measurement case are presented. The solution is entirely parametric as all the information coming from the prior distribution and the measurement are explicit parameters of the solution.} With the proposed approach, the EDH filter can be executed significantly faster than by numeric integration.

The structure of the paper is the following. Section \ref{sec:2} introduces the concept of particle flow filtering along with the exact flow equations and existing filter implementations. General properties of the exact flow equations are discussed in Section \ref{sec:general}. The analytic solution is derived in Section \ref{sec:sol}. Applications of the solution are presented and discussed in Section \ref{sec:app}. Concluding remarks are given in Section \ref{sec:conc}.

\section{The particle flow equations} \label{sec:2}
The particle flow introduced by Daum and Huang originates from the logarithmic form of Bayes' formula. For some state vector $\bb{x}$ and measurement vector $\bb{z}$ we have
\begin{equation} \label{eq:log_Bayes}
    \log p(\bb{x}|\bb{z}) = \log p(\bb{z}|\bb{x}) + \log p(\bb{x}) - \log p(\bb{z}) \, ,
\end{equation}
where $p(\bb{x}|\bb{z})$ is the posterior, $p(\bb{z}|\bb{x})$ is the likelihood, $p(\bb{x})$ is the prior and the normalizing factor is $p(\bb{z})$. By inserting the homotopy parameter $\lambda$ as a coefficient for the likelihood, we arrive at the log-homotopy form of Bayes' formula:
\begin{equation} \label{eq:log_homotopy}
    \log p_{\lambda}(\bb{x}|\bb{z}) = \lambda \log p(\bb{z}|\bb{x}) + \log p(\bb{x}) - \log p_{\lambda}(\bb{z}) \, .
\end{equation}
\hl{The homotopy parameter $\lambda \in [0,1]$ has the initial value 0 and increases to 1. With $\lambda=0$ the right side of \eqref{eq:log_homotopy} gives back the prior and the $\lambda = 1$ case corresponds to \eqref{eq:log_Bayes}.
The Bayesian update is regarded as a continuous transformation of the prior into the posterior, parameterized by the homotopy parameter $\lambda$. In the context of particle filters and by virtue of the fact that samples represent the distributions, the transformation of the density function is manifested as a motion of the particles in the state space}. The dynamics of the motion is modeled by the Itô stochastic differential equation
\begin{equation} \label{eq:sochastic_diff}
    \diff{\bb{X}_{\lambda}} = \bb{f}(\bb{X}_{\lambda}, \lambda) \diff{\lambda} + \boldsymbol{\sigma}(\bb{X}_{\lambda}, \lambda) \diff{\bb{W}_{\lambda}} \, ,
\end{equation}
where $\bb{X}_\lambda$ is a random vector variable, $\bb{W}_\lambda$ is a Wiener process, $\bb{f}$ is the drift and $\boldsymbol{\sigma}$ is the diffusion coefficient. The evolution of the probability density $p$ of the stochastic variable $\bb{X}_\lambda$ is described by the Fokker–Planck equation \cite{risken1996fokker}
\begin{equation} \label{eq:FPE}
    \frac{\partial p}{\partial \lambda} = -\sum_i\frac{\partial}{\partial x_i}p f_i + \frac{1}{2}  \sum_i \sum_j \frac{\partial^2}{\partial x_i \partial x_j} p D_{ij} \, ,
\end{equation}
where $\bb{D} = \boldsymbol{\sigma}\boldsymbol{\sigma}^\top$. In general $p$, $\bb{f}$, and $\bb{D}$ depend on $\lambda$ and $\bb{x}$.

The strategy of the Daum–Huang particle flow filter is the following. The particle update step is achieved by solving the stochastic differential equation \eqref{eq:sochastic_diff} in the interval $\lambda \in [0,1]$ for every particle. The drift vector $\bb{f}$ and diffusion matrix $\boldsymbol{\sigma}$ come from the Fokker–Planck equation. This is an unusual approach since what we have are the boundary conditions and constraints from \eqref{eq:log_homotopy} and what we are looking for are the driving forces. \hl{Creating a specific particle flow is equivalent to the mathematical task of finding a solution for $\bb{f}$ and $\boldsymbol{\sigma}$ based on \eqref{eq:log_homotopy} and \eqref{eq:FPE}.
There is no unique solution to this problem in general, but once we choose one \eqref{eq:sochastic_diff} can be integrated. Further discussion and visualization of how log-homotopy particle flow filters work can be found in \cite{choi2011discussion, crouse2020consideration, daum2011hollywood}.}

Numerous flows have been derived so far \cite{daum2015baker}, \cite{daum2016seven}. The main difference is whether the matrix $\bb{D}$ is neglected or not, or in other words, is the flow deterministic or stochastic.

In this work, the exact flow is considered, which is a deterministic flow with the following assumptions. The drift function has the linear form
\begin{equation} \label{eq:drift}
    \bb{f}(\bb{x}, \lambda) = \bb{A}(\lambda)\bb{x}(\lambda) + \bb{b}(\lambda) \, ,
\end{equation}
and the probability density of $\bb{x}$ is assumed to be Gaussian. Using $\bb{f}$ from \eqref{eq:drift} and neglecting the diffusion, the stochastic equation \eqref{eq:sochastic_diff} can be cast into an ordinary differential equation
\begin{equation} \label{eq:EDHflow}
    \frac{\diff{\bb{x}}(\lambda)}{\diff{\lambda}} = \bb{A}(\lambda)\bb{x}(\lambda) + \bb{b}(\lambda) \, .
\end{equation}

\hl{The exact flow has been introduced in \cite{daum2010exact} and its derivation comes in different flavoures, e.g. \cite{crouse2020consideration, dai2021new}. The matrix $\bb{A}(\lambda)$ and vector $\bb{b}(\lambda)$ have the following forms:}
\begin{align}
     & \bb{A}(\lambda) = -\frac{1}{2}\bb{P} \bb{H}^\top\left(\lambda \bb{H} \bb{P} \bb{H}^\top + \bb{R}\right)^{-1}\bb{H}, \label{eq:A}                                        \\
     & \bb{b}(\lambda) = \left(\bb{I}+2\lambda\bb{A}\right)\left(\left(\bb{I}+\lambda\bb{A}\right)\bb{P}\bb{H}^\top\bb{R}^{-1}\bb{z} + \bb{A}\overline{\bb{x}}\right). \label{eq:b}
\end{align}
The matrices $\bb{P}$, $\bb{H}$, and $\bb{R}$ are the usual ingredients of a Kalman filter and will be discussed in the next section. $\bb{z}$ is the measurement vector, and $\overline{\bb{x}}$ denotes the mean of the predicted Gaussian. The usual assumption is that $\bb{P}$ and $\bb{R}$ are positive definite matrices \cite{daum2010exact} and it can be shown that matrix $\bb{A}$ is stable \cite{dai2021stability}.
\subsection{State estimation with the exact flow}
\begin{figure}
    \centering
    \includegraphics[width=0.6\linewidth]{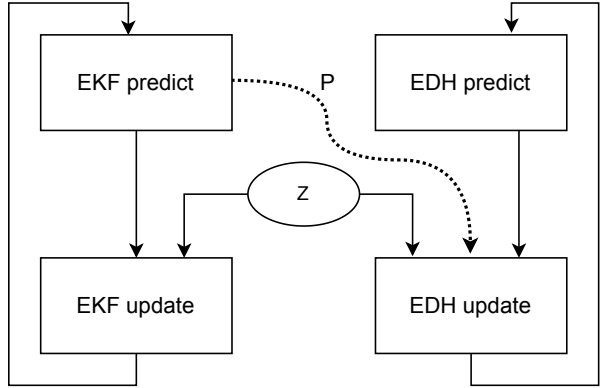}
    \caption{Structure of the parallel running filters. The Extended Kalman filter provides the prediction error covariance matrix for the Exact Daum–Huang particle flow filter.}
    \label{fig:filter}
\end{figure}

As can be seen from equations \eqref{eq:A} and \eqref{eq:b}, the exact Daum–Huang (EDH) flow needs the covariance matrix $\bb{P}$ of the predicted state. For this reason, the exact flow particle filter either needs a parallel EKF (see Fig. \ref{fig:filter}) or a similar filter \cite{choi2011discussion} to provide the matrix $\bb{P}$ or it can be computed from the particle ensemble as the sample covariance. The latter may or may not provide a better performance, depending on the number of particles. In this work, the EKF prediction for $\bb{P}$ will be used.

\hl{Consider an estimation problem with state vector $\bb{x}$, measurement vector $\bb{z}$ and with nonlinear system equations in the form}
\begin{align}
    \bb{x}_{k+1} &= \bb{g}_k(\bb{x}_k) + \bb{w}_k \\
    \bb{z}_k &= \bb{h}_k(\bb{x}_k) + \bb{v}_k \, ,
\end{align}
where $\bb{w}$ and $\bb{v}$ are additive white Gaussian noises (AWGN) with $\Cov(\bb{w}_k)=\bb{Q}_k$ and $\Cov(\bb{v}_k)=\bb{R}_k$, $\bb{g}$ and $\bb{h}$ are the nonlinear state transition and measurement functions, and $k$ is the discrete time index. The estimation of the state vector $\hat{\bb{x}}$ based on the measurement $\bb{z}$ according to an EKF is
\begin{align}
    \hat{\bb{x}}_{k|k-1} &= \bb{g}_k(\hat{\bb{x}}_{k-1|k-1}) \label{eq:EKFpredx} \\
    \bb{P}_{k|k-1} &= \bb{G}_k \bb{P}_{k-1|k-1} \bb{G}_k + \bb{Q}_k \label{eq:EKFpredP} \\
    \bb{S}_k &= \bb{H}_k \bb{P}_{k|k-1} \bb{H}_k^\top + \bb{R}_k \label{eq:EKFres}\\
    \bb{K}_k &= \bb{P}_{k|k-1} \bb{H}_k^\top \bb{S}_k^{-1} \\
    \hat{\bb{x}}_{k|k} &= \hat{\bb{x}}_{k|k-1} + \bb{K}_k\left( \bb{z}_k - \bb{h}_k(\hat{\bb{x}}_{k|k-1})  \right) \\
    \bb{P}_{k|k} &= \left( \bb{I} - \bb{K}_k \bb{H}_k \right) \bb{P}_{k|k-1} \label{eq:EKFupd}\, ,
\end{align}
where the Jacobians $\bb{F}_k$ and $\bb{H}_k$  are
\begin{align}
    \bb{G}_k &= \left. \frac{\partial \bb{g}_k}{\partial \bb{x}} \right|_{\hat{\bb{x}}_{k-1|k-1}} \\
    \bb{H}_k &= \left. \frac{\partial \bb{h}_k}{\partial \bb{x}} \right|_{\hat{\bb{x}}_{k|k-1}} \, . \label{eq:EKF_H}
\end{align}

\hl{For the EDH filter, the prediction is the standard procedure used for ordinary particle filters, that is the propagation of the particles through the motion model. Every particle $\bb{x}^{i}$ is drawn form the prior distribution: }
\begin{equation} \label{eq:particlePred}
    \bb{x}_{k|k-1}^{i} \sim \mathcal{N}\left( \bb{g}_k(\bb{x}_{k-1|k-1}^{i}), \bb{Q}_k \right) \, ,
\end{equation}
where $\mathcal{N}(\bb{m}, \bb{C})$ represents a Gaussian distribution with mean $\bb{m}$ and covariance $\bb{C}$.

To update a particle the EDH flow equation \eqref{eq:EDHflow} is used. \hl{The initial value of the flow equation is a particle $\bb{x}_{0}^{i}$ coming from the prior distribution thus it belongs to $\lambda = 0$:}
\begin{equation}
    \bb{x}_{0}^{i} \defeq \bb{x}^{i}(\lambda = 0) = \bb{x}_{k|k-1}^{i} \, .
\end{equation}
\hl{To get the posterior distribution we need to solve \eqref{eq:EDHflow} for $\bb{x}_{N}^{i} \defeq \bb{x}^{i}(\lambda = 1)$. In practice it means an $N$ step Euler integration in the form}
\begin{equation} \label{eq:EDH_upd}
    \bb{x}_n^{i} = \bb{x}_{n-1} + \bb{f}(\overline{\bb{x}}_{n-1}^{i}, \lambda) \Delta \lambda \, \, \, \, \,  (n = 1 \dots N) \, ,
\end{equation}
where $\Delta \lambda = 1/N$ is the step size. The mean value $\overline{\bb{x}}_{n-1}^{i}$ is computed for the particle set for which the linearization in \eqref{eq:EKF_H} happens. If the linearization happens for every particle, thus $\bb{f}(\bb{x}_{n-1}^{i}, \lambda)$ is used instead of $\bb{f}(\overline{\bb{x}}_{n-1}^{i}, \lambda)$ then more accuracy can be achieved for the cost of greater computation. This approach is referred to as the localized exact Daum–Huang (LEDH) filter \cite{ding2012implementation}. The posterior estimate is given by the particle ensemble $\bb{x}_N^{i}$. Note that particle weights or resampling are not part of the algorithm, and the matrix $\bb{P}$ for the drift function $\bb{f}$ comes from \eqref{eq:EKFpredP}.

In \cite{khan2014non} three variants of the EDH filter were compared: the original, the LEDH, and a coupled, where the EDH output is fed back to the EKF.

Several papers reported particle flow techniques inserted into conventional particle filtering frameworks to create a proposal distribution \cite{li2015particle, li2017particle}. Clustering the particle set can decrease the computational requirement or increase the performance of the EDH filter \cite{li2016fast, orenbacs2019clustered}.

Random finite set (RFS) approaches to state estimation and multi-object tracking \cite{mahler2014advances} also needs particle filter implementations due to lack of analytic solutions \cite{ristic2013particle}. RFS-based state estimators were combined with particle flow methods such as the multi-object probability hypothesis density \cite{zhao2016gaussian} or the labeled multi-Bernoulli filter \cite{saucan2017particle}.
Gaussian mixture models can also be used for multi-object particle flow estimators \cite{pal2017gaussian}.

\section{General properties of the exact flow equation} \label{sec:general}
The system of differential equations representing the flow is
\begin{equation} \label{eq:theEQ}
    \frac{\diff{}\bb{x}(\lambda)}{\diff \lambda} = \bb{A}(\lambda) \bb{x}(\lambda) + \bb{b}(\lambda)
\end{equation}
with dimensions $\bb{x}, \bb{b} \in \mathbb{R}^{n_x}, \bb{A} \in \mathbb{R}^{n_x\times n_x}$, and $\lambda \in [0,1]$. This is a first-order linear time-varying inhomogeneous ordinary matrix differential equation. A closed form  solution can be found if the matrix $\bb{A}(\lambda)$ is commutative \cite[Chapter 7]{DiffEq1982}. In particular, if $\bb{A}(\lambda)$ commutes with its integral in the domain of interest, that is
\begin{equation} \label{eq:theCommutator}
    \bb{A}(\lambda) \int_{0}^\lambda \bb{A}(\tau) \mathrm{d}\tau - \int_{0}^\lambda \bb{A}(\tau) \mathrm{d}\tau \, \bb{A}(\lambda) = 0 \, ,
\end{equation}
the state transition matrix can be constructed as
\begin{equation}
    \bb{\Phi}(\lambda,0) = \exp\left(\int_{0}^\lambda\bb{A}(\tau)\mathrm{d}\tau\right) \, ,
\end{equation}
and the general solution to \eqref{eq:theEQ} is
\begin{equation} \label{eq:theSolution}
    \bb{x}(\lambda) = \bb{\Phi}(\lambda,0)\bb{x}(0) + \int_{0}^\lambda \bb{\Phi}(\lambda,\tau) \bb{b}(\tau) \mathrm{d}\tau \, ,
\end{equation}
where $\bb{x}(0) = \bb{x}_0$ is the initial condition.
The requirements towards $\bb{A}(\lambda)$ is that it is at least piecewise continuous.

The matrices in $\bb{A}(\lambda)$ and $\bb{b}(\lambda)$ are all real-valued with the following properties. $\bb{P}$ and $\bb{R}$ are symmetric positive definite of size $n_x\times n_x$ and $n_z \times n_z$. $\bb{H}$ is of size $n_z \times n_x$. $\bb{z}$ and $\bar{\bb{x}}$ are column vectors with $n_z$ and $n_x$ elements. From practical considerations, an additional assumption can be made. $\bb{H}$ is the measurement matrix which is full rank; otherwise, it would mean we measure the same quantity more than once.

To apply the solution \eqref{eq:theSolution} it needs to be verified that the commutator in \eqref{eq:theCommutator} equals zero. In particular we will show that $\bb{A}(\lambda)$ and $\bb{A}(\tau)$ commute for every $\lambda, \tau \in [0,1]$.

The following property will be used in the proof. If ${(\bb{M}+\bb{I})^{-1}}$ exists for some square matrix $\bb{M}$ then from the identity
\begin{equation}
    (\bb{M} +\bb{I})^{-1} =  \bb{I} - (\bb{M} +\bb{I})^{-1}\bb{M} = \bb{I}-\bb{M}(\bb{M} +\bb{I})^{-1}
\end{equation}
it follows that $\bb{M}$ commutes with $(\bb{M} +\bb{I})^{-1}$ and also with $(\lambda \bb{M} +\bb{I})^{-1}$. Now we need to bring $\bb{A}(\lambda)$ to a form such that the above property can be exploited.
The product ${\bb{A}(\lambda)\bb{A}(\tau)}$, without the scalar coefficient, has the form
\begin{equation}
    \bb{P} \bb{H}^\top\!\!\left(\lambda \bb{H} \bb{P} \bb{H}^\top \!\!+\! \bb{R}\right)^{-1}\!\!\bb{H} \bb{P} \bb{H}^\top\!\!\left(\tau \bb{H} \bb{P} \bb{H}^\top \!\!+\! \bb{R}\right)^{-1}\!\!\bb{H}.
\end{equation}
We can pull out $\bb{R}$ from the inverses as
\begin{equation}
    \bb{P} \bb{H}^\top\bb{R}^{-1}\left(\lambda \bb{H} \bb{P} \bb{H}^\top\bb{R}^{-1} + \bb{I}\right)^{-1} 
    \bb{H} \bb{P} \bb{H}^\top\bb{R}^{-1}\left(\tau \bb{H} \bb{P} \bb{H}^\top\bb{R}^{-1} + \bb{I}\right)^{-1}\bb{H}
\end{equation}
and push the factor ${\bb{H} \bb{P} \bb{H}^\top\bb{R}^{-1}}$ through the second inverse to the right:
\begin{equation}
    \bb{P} \bb{H}^\top\bb{R}^{-1}\left(\lambda \bb{H} \bb{P} \bb{H}^\top\bb{R}^{-1} + \bb{I}\right)^{-1} 
    \left(\tau \bb{H} \bb{P} \bb{H}^\top\bb{R}^{-1} + \bb{I}\right)^{-1}\bb{H} \bb{P} \bb{H}^\top\bb{R}^{-1}\bb{H} \, .
\end{equation}
From the product of the two inverses, we form the expression
\begin{equation}
    \left(\left(\tau \bb{H} \bb{P} \bb{H}^\top\bb{R}^{-1} + \bb{I}\right)\left(\lambda \bb{H} \bb{P} \bb{H}^\top\bb{R}^{-1} + \bb{I}\right) \right)^{-1} \, ,
\end{equation}
which, after expanding into
\begin{equation}
        \left( \tau \lambda (\bb{H} \bb{P} \bb{H}^\top\bb{R}^{-1})^2 + (\tau+\lambda) \bb{H} \bb{P} \bb{H}^\top\bb{R}^{-1} + \bb{I} \right)^{-1} \, ,
\end{equation}
can be seen as symmetric in $\lambda$ and $\tau$. This concludes the proof, that $\bb{A}(\lambda)\bb{A}(\tau) = \bb{A}(\tau)\bb{A}(\lambda)$, thus the matrix $A(\lambda)$ is commutative on $\lambda \in [0,1]$.

\section{Solution for the scalar measurement case} \label{sec:sol}
From $n_z = 1$ it follows that $\bb{H}$ is a row vector of size $n_z$, $\bb{R} \in \mathbb{R}^+$ and ${\bb{HPH}^\top}$ is also a scalar.

To construct the solution we need the state transition matrix $\bb{\Phi}(\lambda,\lambda_0)$. Using
\begin{equation}
    \int_{\lambda_0}^\lambda \!\!\bb{A}(\tau) \mathrm{d} \tau = -\frac{1}{2}\bb{\bb{P}} \bb{H}^\top \!\int_{\lambda_0}^\lambda \!\!\left( \tau \bb{H} \bb{P} \bb{H}^\top \!+ \bb{R} \right)^{-1} \mathrm{d} \tau \bb{H}
\end{equation}
and
\begin{align}
     & \int_{\lambda_0}^\lambda\!\! \left( \tau \bb{H} \bb{P} \bb{H}^\top \!\!+\! \bb{R} \right)^{-1} \mathrm{d} \tau = \left[\frac{\log\left( \tau \bb{H} \bb{P} \bb{H}^\top \!\!+\! \bb{R} \right)}{\bb{H} \bb{P} \bb{H}^\top}\right]_{\lambda_0}^\lambda
\end{align}
we arrive at
\begin{equation}
    \int_{\lambda_0}^\lambda\!\! \bb{A}(\tau) \mathrm{d} \tau = -\frac{1}{2}\frac{\bb{P} \bb{H}^\top}{\bb{H} \bb{P} \bb{H}^\top}\log\left( \frac{\lambda \bb{H} \bb{P} \bb{H}^\top + \bb{R}}{\lambda_0 \bb{H} \bb{P} \bb{H}^\top + \bb{R}} \right) \bb{H}.
\end{equation}
The denominator $\lambda_0 \bb{H} \bb{P} \bb{H}^\top + \bb{R}$ cannot take zero value because $\bb{P}$ being a positive definite matrix ${\bb{H} \bb{P} \bb{H}^\top}$ is always positive.

Collecting the scalar terms in the factor ${\beta(\lambda,\lambda_0)}$ the state transition matrix gains the form
\begin{align}
    \bb{\Phi}(\lambda,\lambda_0) = \exp{\left(\beta(\lambda, \lambda_0) \bb{P} \bb{H}^\top \bb{H}\right)}
\end{align}
where
\begin{align} \label{eq:beta}
    \beta(\lambda,\lambda_0) & = \log\left(\frac{\lambda \bb{H} \bb{P} \bb{H}^\top + \bb{R}}{\lambda_0 \bb{H} \bb{P} \bb{H}^\top + \bb{R}}\right)^{-1/(2\bb{H} \bb{P} \bb{H}^\top)}  
\end{align}
The matrix product $\bb{P}\bb{H}^\top \bb{H}$ has rank 1 due to the fact that $\bb{H}^\top \bb{H}$ is rank 1 and $\bb{P}$ is positive definite.

The exponential of a rank 1 square matrix $\beta \bb{M}, (\beta \in \mathbb{R})$ can be expressed as \cite{bernstein1993some}
\begin{equation}
    \mathrm{e}^{\beta \bb{M}} = \bb{I} + \bb{M}(\mathrm{e}^{\beta \tr(\bb{M})}-1)/ \tr(\bb{M}) \, .
    \label{eq:rank1_exp}
\end{equation}
Using \eqref{eq:rank1_exp} the state transition matrix is provided by
\begin{equation} \label{eq:FI1}
     \bb{\Phi}(\lambda,\lambda_0) = \bb{I} +  \frac{\bb{P} \bb{H}^\top \bb{H}}{\tr(\bb{P} \bb{H}^\top \bb{H})} \left( \mathrm{e}^{\beta(\lambda, \lambda_0) \tr(\bb{P} \bb{H}^\top \bb{H})} - 1\right) \, .
\end{equation}
The product $\bb{P} \bb{H}^\top \bb{H}$ in the trace can be permuted and since $\bb{H} \bb{P} \bb{H}^\top$ is a scalar
\begin{equation} \label{eq:trace}
    \tr(\bb{P} \bb{H}^\top \bb{H}) = \bb{H} \bb{P} \bb{H}^\top
\end{equation}
follows.

By substituting $\beta(\lambda, \lambda_0)$ from \eqref{eq:beta} into \eqref{eq:FI1} we observe that $\tr(\bb{P} \bb{H}^\top \bb{H})$ goes up to the exponent in the logarithm, and due to \eqref{eq:trace} gets cancelled. After that, the exp and the log can cancel each other, yielding
\begin{equation} \label{eq:FI2}
     \bb{\Phi}(\lambda,\lambda_0) = \bb{I} +  \frac{\bb{P} \bb{H}^\top \bb{H}}{\bb{H} \bb{P} \bb{H}^\top} \bigg( 
     \left(\frac{\lambda \bb{H} \bb{P} \bb{H}^\top + \bb{R}}{\lambda_0 \bb{H} \bb{P} \bb{H}^\top + \bb{R}}\right)^{\nicefrac{-1}{2}}
     \!\!\!\!-\! 1\! \bigg) \, .
\end{equation}

From now on, the following abbreviations will be used:
\begin{align}
    \bb{M} & \defeq \bb{P} \bb{H}^\top \bb{H} \\
    k(\lambda) &\defeq \lambda \bb{H} \bb{P} \bb{H}^\top + \bb{R} \\
    \bb{w} &\defeq \bb{P} \bb{H}^\top \bb{R}^{-1} \bb{z} \\
    p &\defeq \bb{H} \bb{P} \bb{H}^\top \\
    r &\defeq \bb{R} \, ,
\end{align}
which enables writing $\bb{\Phi}(\lambda,\lambda_0)$ in the form
\begin{equation} \label{eq:Phi}
    \bb{\Phi}(\lambda,\lambda_0) = \bb{I} + \frac{\bb{M}}{p} \left(  k(\lambda_0)^{\nicefrac{1}{2}}k(\lambda)^{\nicefrac{-1}{2}} - 1 \right) \, .
\end{equation}

Inspecting the matrix $\bb{H}^\top \bb{H}$ it can be seen that it is of rank 1 and has eigenvalues $0$ and $\bb{H}\bb{H}^\top$ and eigenvector $\bb{H}^\top$, thus the eigenspace of the matrix $\bb{H}^\top \bb{H}$ is in the direction of $\bb{H}^\top$. 
The matrix $\bb{M}=\bb{P}\bb{H}^\top\bb{H}$ is also of rank 1 and because $\bb{H}^\top\bb{H}$ transforms vectors in the direction of $\bb{H}^\top$, $\bb{P}\bb{H}^\top\bb{H}$ has eigenspace in the direction of $\bb{P}\bb{H}^\top$. Now we have
\begin{equation}
    \bb{P} \bb{H}^\top \bb{H} \bb{P} \bb{H}^\top = \alpha \bb{P} \bb{H}^\top
\end{equation}
therefore $\alpha = \bb{H} \bb{P} \bb{H}^\top$ is an eigenvalue, $\bb{P} \bb{H}^\top$ and $\bb{w}=\bb{P}\bb{H}^\top \bb{R}^{-1}z$ is an eigenvector of $\bb{M}$. From this follows that
\begin{equation} \label{eq:Mwpw}
    \bb{M}\bb{w} = \bb{H}\bb{P}\bb{H}^\top \bb{w} = p\bb{w} \, .
\end{equation}
\subsection{Inhomogeneous part}
The inhomogeneous part of the solution is
\begin{equation}
    \int_{\lambda_0}^\lambda \bb{\Phi}(\lambda,\tau) \bb{b}(\tau)  \diff{\tau}
\end{equation}
where $\bb{b}(\tau)$ from \eqref{eq:b} in expanded form is
\begin{multline}
    \bb{b}(\tau) = \bb{P} \bb{H}^\top \bb{R}^{-1} \bb{z} + \bb{A}(\tau) \bar{\bb{x}} + 3 \tau \bb{A}(\tau) \bb{P} \bb{H}^\top \bb{R}^{-1} \bb{z} \\ + 2 \tau \bb{A}(\tau)^2 \bar{\bb{x}} + 2 \tau^2 \bb{A}(\tau)^2 \bb{P} \bb{H}^\top \bb{R}^{-1} \bb{z} \, ,
\end{multline}
which will be handled as a sum of five terms: $\bb{b} = \bb{b}_0+\bb{b}_1+\bb{b}_2+\bb{b}_3+\bb{b}_4$.

Using
\begin{align}
    \bb{A}(\tau) & = -\frac{1}{2}\bb{P} \bb{H}^\top\left(\tau \bb{H} \bb{P} \bb{H}^\top + \bb{R}\right)^{-1}\bb{H} \nonumber \\
    &= -\frac{1}{2} \bb{M} k(\tau)^{-1}
\end{align}
and
\begin{align}
    \bb{A}(\tau)^2 &= \frac{1}{4} \left( \tau \bb{H} \bb{P} \bb{H}^\top + \bb{R} \right)^{-2} \bb{P} \bb{H}^\top \bb{H} \bb{P} \bb{H}^\top \bb{H} \nonumber \\
    & = \frac{1}{4} \bb{M} pk(\tau)^{-2}
\end{align}
we have
\begin{align}
     \bb{b}_0 &= \bb{P} \bb{H}^\top \bb{R}^{-1} \bb{z} = \bb{w}\\
    \bb{b}_1 & = \bb{A} \bar{\bb{x}} \nonumber \\
    &= -\frac{1}{2}\bb{PH}^\top (\tau \bb{H}\bb{P} \bb{H}^\top + \bb{R})^{-1} \bb{H} \bar{\bb{x}} \nonumber \\ 
    &= -\frac{1}{2} \bb{M} \bar{\bb{x}} k(\tau)^{-1} \\
    \bb{b}_2 &= 3 \tau \bb{A} \bb{P} \bb{H}^\top \bb{R}^{-1} \bb{z} \nonumber \\
    &= -\frac{3}{2} \tau \bb{H} \bb{P} \bb{H}^\top (\tau \bb{H} \bb{P} \bb{H}^\top + \bb{R})^{-1} \bb{P} \bb{H}^\top \bb{R}^{-1} \bb{z} \nonumber\\
    &= -\frac{3}{2} \bb{w} p \tau k(\tau)^{-1} \\
    \bb{b}_3 &= 2 \tau \bb{A}^2 \bar{\bb{x}} \nonumber \\
    &= \frac{1}{2} \tau \bb{H} \bb{P} \bb{H}^\top (\tau \bb{H} \bb{P} \bb{H}^\top + \bb{R})^{-2} \bb{P} \bb{H}^\top \bb{H} \bar{\bb{x}} \nonumber \\
    &= \frac{1}{2} \bb{M} \bar{\bb{x}} \tau p k(\tau)^{-2} \\
    \bb{b}_4 &= 2 \tau^2 \bb{A}^2 \bb{P} \bb{H}^\top \bb{R}^{-1} \bb{z} \nonumber \\
    &= \frac{1}{2}\tau^2 (\bb{H} \bb{P} \bb{H}^\top)^2(\tau \bb{H} \bb{P} \bb{H}^\top + \bb{R})^{-2} \bb{P} \bb{H}^\top \bb{R}^{-1} \bb{z} \nonumber \\
    &= \frac{1}{2} \bb{w} \tau^2 p^2k(\tau)^{-2}
\end{align}

In the next subsections the five integrals of the form
\begin{equation}
    \int_{\lambda_0}^\lambda \bb{\Phi}(\lambda,\tau)\bb{b}_i(\tau) \diff{\tau} \text{ $(i=0\dots 4)$}
\end{equation}
 will be evaluated.
 
\subsubsection{The $b_0$ term}
The integrand is
\begin{equation} \label{eq:FIb0}
    \bb{\Phi}(\lambda, \tau)\bb{b}_0 = \bb{w} + \frac{\bb{M}\bb{w}}{p}\left( k(\tau)^{\nicefrac{1}{2}}k(\lambda)^{\nicefrac{-1}{2}} \right) - \frac{\bb{M}\bb{w}}{p}
\end{equation}
and since $\bb{M}\bb{w} = p \bb{w}$ \eqref{eq:FIb0} reduces to
\begin{equation}
    \bb{\Phi}(\lambda, \tau)\bb{b}_0 = \bb{w}\left( k(\tau)^{\nicefrac{1}{2}}k(\lambda)^{\nicefrac{-1}{2}}\right) \, .
\end{equation}
The integral has the form
\begin{equation}
    \bb{\Psi}_0(\lambda,\lambda_0) = \int_{\lambda_0}^{\lambda}  \bb{\Phi}(\lambda, \tau)\bb{b}_0 \diff{\tau}= \bb{w} k(\lambda)^{\nicefrac{-1}{2}} \int_{\lambda_0}^{\lambda} k(\tau)^{\nicefrac{1}{2}} \diff{\tau} .
\end{equation}
Using
\begin{align}
    \int_{\lambda_0}^{\lambda} (\tau p+r)^{\nicefrac{1}{2}} \diff{\tau} &=\left[ \frac{2}{3p} (\tau p+r)^{\nicefrac{3}{2}} \right]_{\lambda_0}^{\lambda} =\frac{2}{3p}\left(k(\lambda)^{\nicefrac{3}{2}} - k(\lambda_0)^{\nicefrac{3}{2}}\right)
\end{align}
we arrive at
\begin{equation} \label{eq:b0}
    \bb{\Psi}_0(\lambda,\lambda_0) = \frac{2}{3}\frac{\bb{w}}{p}\left( k(\lambda) - k(\lambda_0)^{\nicefrac{3}{2}} k(\lambda)^{\nicefrac{-1}{2}} \right) \, .
\end{equation}

\subsubsection{The $b_1$ term}

The integrand is
\begin{align}
\begin{split}
    &\bb{\Phi}(\lambda, \tau) \bb{b}_1  = -\frac{1}{2}\left( \!\bb{I} \!+\! \frac{\bb{M}}{p} \left(k(\tau)^{\nicefrac{1}{2}} k(\lambda)^{\nicefrac{-1}{2}} \!- \!1 \right)\! \right)\! \bb{M}\bar{\bb{x}} k(\tau)^{-1} \\
\end{split} \nonumber \\
&=-\frac{\bb{M}\bar{\bb{x}}}{2} k(\tau)^{-1} - \frac{\bb{M}^2\bar{\bb{x}}}{2p}k(\tau)^{\nicefrac{-1}{2}}k(\lambda)^{\nicefrac{-1}{2}} + \frac{\bb{M}^2\bar{\bb{x}}}{2p}k(\tau)^{-1}
\end{align}
Since $\bb{M}$ is a rank 1 matrix $\bb{M}^2 =\tr(\bb{M}) \bb{M} = p \bb{M}$ follows, thus the first and the last terms cancel and only $k(\tau)^{\nicefrac{-1}{2}}$ needs to be integrated:
\begin{align}
    \bb{\Psi}_1(\lambda,\lambda_0) &= -\frac{\bb{M}\bar{\bb{x}}}{2} k(\lambda)^{\nicefrac{-1}{2}} \int_{\lambda_0}^{\lambda} k(\tau)^{\nicefrac{-1}{2}} \diff{\tau} 
\end{align}
Using
\begin{align}
    \int_{\lambda_0}^{\lambda} (\tau p+r)^{\nicefrac{-1}{2}} \diff{\tau} &= \left[\frac{2}{p} (\tau p+r)^{\nicefrac{1}{2}}\right]_{\lambda_0}^{\lambda} =\frac{2}{p}\left(k(\lambda)^{\nicefrac{1}{2}}-k(\lambda_0)^{\nicefrac{1}{2}}\right)
\end{align}
we have
\begin{align}
    \bb{\Psi}_1(\lambda,\lambda_0) &= -\frac{1}{2}\bb{M}\bar{\bb{x}}k(\lambda)^{\nicefrac{-1}{2}}\frac{2}{p}\left( k(\lambda)^{\nicefrac{1}{2}} - k(\lambda_0)^{\nicefrac{1}{2}} \right) \nonumber \\
     &= \frac{1}{p}\bb{M}\bar{\bb{x}}\left( k(\lambda)^{\nicefrac{-1}{2}}k(\lambda_0)^{\nicefrac{1}{2}} - 1 \right) \label{eq:b1}
\end{align}

\subsubsection{The $b_2$ term}
The integrand is
\begin{align}
\begin{split}
    &\bb{\Phi}(\lambda, \tau) \bb{b}_2  = -\frac{3}{2}\left( \!\bb{I} \!+\! \frac{\bb{M}}{p} \left(k(\tau)^{\nicefrac{1}{2}} k(\lambda)^{\nicefrac{-1}{2}} \!- \!1 \right)\! \right)\! \bb{w} p \tau k(\tau)^{-1} \\
\end{split} \nonumber \\
&=-\frac{3}{2}\tau\left(\bb{w} p k(\tau)^{-1} + \bb{M}\bb{w} k(\tau)^{\nicefrac{-1}{2}}k(\lambda)^{\nicefrac{-1}{2}} - \bb{M}\bb{w} k(\tau)^{-1}\right)
\end{align}
Due to \eqref{eq:Mwpw} the first and last terms in the parenthesis cancel and
\begin{equation}
    \bb{\Phi}(\lambda, \tau) \bb{b}_2 = -\frac{3}{2}p\bb{w} \tau k(\tau)^{\nicefrac{-1}{2}} k(\lambda)^{\nicefrac{-1}{2}}
\end{equation}
remains to be integrated.
The integral is
\begin{align}
    \bb{\Psi}_2(\lambda,\lambda_0) &= \int_{\lambda_0}^{\lambda} \bb{\Phi}(\lambda, \tau) \bb{b}_2 \diff{\tau}   = -\frac{3}{2}p\bb{w} k(\lambda)^{\nicefrac{-1}{2}}  \int_{\lambda_0}^{\lambda} \tau k(\tau)^{\nicefrac{-1}{2}} \diff{\tau}
\end{align}
Using
\begin{align}
    \int_{\lambda_0}^{\lambda} \!\tau(p \tau + r&)^{\nicefrac{-1}{2}} \diff{\tau}\!=\! \left[\frac{2\tau(p\tau+r)^{\nicefrac{1}{2}}}{3p}\!-\!\frac{4r(p\tau+r)^{\nicefrac{1}{2}}}{3p^2}\right]_{\lambda_0}^{\lambda} \nonumber \\
    &=\left[\frac{2\tau k(\tau)^{\nicefrac{1}{2}}}{3p}-\frac{4rk(\tau)^{\nicefrac{1}{2}}}{3p^2}\right]_{\lambda_0}^{\lambda}  \nonumber \\
    &= \frac{2k(\lambda)^{\nicefrac{1}{2}}}{3p^2}(p\lambda-2r) - \frac{2k(\lambda_0)^{\nicefrac{1}{2}}}{3p^2}(p\lambda_0-2r)
\end{align}
we arrive at
\begin{align}
\begin{split}
     \bb{\Psi}_2(\lambda,\lambda_0) =& -\frac{3}{2}p\bb{w}k(\lambda)^{\nicefrac{-1}{2}}\frac{2}{3p^2}  \left(k(\lambda)^{\nicefrac{1}{2}}(p\lambda-2r) - k(\lambda_0)^{\nicefrac{1}{2}}(p\lambda_0-2r)\right)
     \end{split} \nonumber \\
     =& \frac{\bb{w}}{p}\left( 2r-p\lambda + k(\lambda)^{\nicefrac{-1}{2}}k(\lambda_0)^{\nicefrac{1}{2}}(2r-p\lambda_0)\right) \label{eq:b2}
\end{align}

\subsubsection{The $b_3$ term}
The integrand is
\begin{align}
\begin{split}
    \bb{\Phi}(\lambda, \tau) \bb{b}_3  &= 
    \frac{1}{2} \left( \!\bb{I} \!+\! \frac{\bb{M}}{p} \left(k(\tau)^{\nicefrac{1}{2}} k(\lambda)^{\nicefrac{-1}{2}} \!- \!1 \right)\! \right) \bb{M} \bar{\bb{x}} \tau p k(\tau)^{-2}
\end{split} \nonumber \\
\begin{split}
     &= \frac{\bb{M}\bar{\bb{x}}}{2}p\tau  k(\tau)^{-2} \!+\! \frac{\bb{M}^2\bar{\bb{x}}}{2}\tau k(\tau)^{\nicefrac{-3}{2}}k(\lambda)^{\nicefrac{-1}{2}} \!-\! \frac{\bb{M}^2\bar{\bb{x}}}{2}\tau k(\tau)^{-2}
\end{split}
\end{align}
Again, using $\bb{M}^2 = p \bb{M}$ the first and last terms cancel thus
\begin{equation}
     \bb{\Phi}(\lambda, \tau) \bb{b}_3  = \frac{\bb{M}\bar{\bb{x}}}{2}p\tau k(\tau)^{\nicefrac{-3}{2}}k(\lambda)^{\nicefrac{-1}{2}} \, .
\end{equation}
Using
\begin{align}
    &\int_{\lambda_0}^{\lambda} \tau (p\tau+r)^{\nicefrac{-3}{2}} \diff{\tau} = \left[\frac{2}{p^2}(p\tau+2r)(p\tau+r)^{\nicefrac{-1}{2}}\right]_{\lambda_0}^{\lambda} \nonumber \\
    &=\frac{2}{p^2}\left((p\lambda+2r)k(\lambda)^{\nicefrac{-1}{2}} - (p\lambda_0+2r)k(\lambda_0)^{\nicefrac{-1}{2}} \right)
\end{align}
we get
\begin{align}
     \bb{\Psi}_3(\lambda,\lambda_0) &= \frac{\bb{M}}{p}\bar{\bb{x}}k(\lambda)^{\nicefrac{-1}{2}}  
     \left(  (p\lambda+2r)k(\lambda)^{\nicefrac{-1}{2}} - (p\lambda_0+2r)k(\lambda_0)^{\nicefrac{-1}{2}}\right) \nonumber \\
     &= \frac{\bb{M}}{p}\bar{\bb{x}}  
     \left(  (p\lambda+2r)k(\lambda)^{-1} - (p\lambda_0+2r)k(\lambda_0)^{\nicefrac{-1}{2}}k(\lambda)^{\nicefrac{-1}{2}}\right) \label{eq:b3}
\end{align}

\subsubsection{The $b_4$ term}
The integrand is
\begin{align}
\begin{split}
    &\bb{\Phi}(\lambda, \tau) \bb{b}_4  = 
    \frac{1}{2}\left( \!\bb{I} \!+\! \frac{\bb{M}}{p} \left(k(\tau)^{\nicefrac{1}{2}} k(\lambda)^{\nicefrac{-1}{2}} \!- \!1 \right)\! \right) \bb{w} \tau^2 p^2k(\tau)^{-2}
\end{split}  \nonumber \\
\begin{split}
     &= \frac{1}{2} \bb{w} \tau^2 p^2k(\tau)^{-2} +  \frac{1}{2p}\bb{M}\bb{w} \tau^2p^2k(\tau)^{\nicefrac{-3}{2}}k(\lambda)^{\nicefrac{-1}{2}} - \frac{\bb{M}\bb{w}}{2p}\tau^2p^2k(\tau)^{-2}
\end{split}
\end{align}
The first and last terms cancel thus the integral reduces to
\begin{equation}
    \bb{\Psi}_4(\lambda,\lambda_0) = \frac{1}{2} p^2\bb{w} k(\lambda)^{\nicefrac{-1}{2}} \int_{\lambda_0}^{\lambda}  \tau^2k(\tau)^{\nicefrac{-3}{2}}\diff{\tau} \, .
\end{equation}
Using
\begin{equation}
    \int_{\lambda_0}^{\lambda} \tau^2(p\tau+r)^{\nicefrac{-3}{2}} \diff{\tau} = \left[ \frac{2(p^2\tau^2-4pr\tau-8r^2)}{3p^2(p\tau+r)^{\nicefrac{1}{2}}} \right]_{\lambda_0}^{\lambda}
\end{equation}
we can write
\begin{multline} \label{eq:b4}
    \bb{\Psi}_4(\lambda,\lambda_0) = \frac{1}{3p}\bb{w} k(\lambda)^{\nicefrac{-1}{2}}
    \left( \frac{p^2\lambda^2-4pr\lambda-8r^2}{(p\lambda+r)^{\nicefrac{1}{2}}}\!\! -\!\! \frac{p^2\lambda_0^2-4pr\lambda_0-8r^2}{(p\lambda_0+r)^{\nicefrac{1}{2}}} \right)
\end{multline}

\subsection{Solution for homotopy}
The general solution for the differential equation \eqref{eq:theEQ} with initial state $\bb{x}(\lambda_0)$ assembles from the homogeneous part \eqref{eq:Phi} and the five inhomogeneous parts \eqref{eq:b0}, \eqref{eq:b1}, \eqref{eq:b2}, \eqref{eq:b3}, and \eqref{eq:b4} as
\begin{align} 
    \bb{x}(\lambda) &= \bb{\Phi}(\lambda, \lambda_0)\bb{x}(\lambda_0) + \bb{\Psi}_0(\lambda,\lambda_0) \nonumber \\
    &+ \bb{\Psi}_1(\lambda,\lambda_0) + \bb{\Psi}_2(\lambda,\lambda_0) + \bb{\Psi}_3(\lambda,\lambda_0) + \bb{\Psi}_4(\lambda,\lambda_0) \, . \label{eq:aedh_sol}
\end{align}
For the homotopy equation, the integration limits are $\lambda_0=0$ and $\lambda = 1$ thus the solution can be simplified to
\begin{align} \label{eq:sol_homFI}
        \bb{\Phi}(1,0) &= \bb{I} + \frac{\bb{M}}{p} \left(  k(0)^{\nicefrac{1}{2}}k(1)^{\nicefrac{-1}{2}} - 1 \right) \nonumber \\
        &= \bb{I} + \frac{\bb{M}}{p} \left(  r^{\nicefrac{1}{2}}(p+r)^{\nicefrac{-1}{2}} - 1 \right) \\
    \bb{\Psi}_0 &= \frac{2}{3}\frac{\bb{w}}{p}\left( k(1) - k(0)^{\nicefrac{3}{2}} k(0)^{\nicefrac{-1}{2}} \right) \nonumber \\
    &= \frac{2}{3}\frac{\bb{w}}{p}\left( p+r - r^{\nicefrac{3}{2}} (p+r)^{\nicefrac{-1}{2}} \right)
\end{align}
\begin{align}
    \bb{\Psi}_1 &= \frac{1}{p}\bb{M}\bar{\bb{x}}\left( k(1)^{\nicefrac{-1}{2}}k(0)^{\nicefrac{1}{2}} - 1 \right)  \nonumber \\
    &=  \frac{1}{p}\bb{M}\bar{\bb{x}}\left( (p+r)^{\nicefrac{-1}{2}}r^{\nicefrac{1}{2}} - 1 \right)
\end{align}
\begin{align}
     \bb{\Psi}_2(\lambda) &= \frac{\bb{w}}{p}\left( 2r-p + 2r k(1)^{\nicefrac{-1}{2}}k(0)^{\nicefrac{1}{2}}\right) \nonumber \\
     &= \frac{\bb{w}}{p}\left( 2r-p + 2 (p+r)^{\nicefrac{-1}{2}}r^{\nicefrac{3}{2}}\right)
\end{align}
\begin{align}
     \bb{\Psi}_3 &= \frac{\bb{M}}{p}\bar{\bb{x}}  \left(  (p+2r)k(1)^{-1} - 2r k(0)^{\nicefrac{-1}{2}}k(1)^{\nicefrac{-1}{2}}\right) \nonumber \\
     &= \frac{\bb{M}}{p}\bar{\bb{x}} \left(  (p+2r)(p+r)^{-1} - 2 r^{\nicefrac{1}{2}}(p+r)^{\nicefrac{-1}{2}}\right)
\end{align}
\begin{align} \label{eq:sol_homb4}
    \bb{\Psi}_4 &= \frac{1}{3p}\bb{w} k(1)^{\nicefrac{-1}{2}}
    \left( \frac{p^2-4pr-8r^2}{(p+r)^{\nicefrac{-1}{2}}} - \frac{-8r^2}{(r)^{\nicefrac{-1}{2}}} \right) \nonumber \\
    &= \frac{1}{3p}\bb{w} (p+r)^{\nicefrac{-1}{2}}
    \left( \frac{p^2-4pr-8r^2}{(p+r)^{\nicefrac{-1}{2}}} - \frac{-8r^2}{(r)^{\nicefrac{-1}{2}}} \right) \nonumber \\
    &= \frac{1}{3p}\bb{w} 
    \left( p^2-4pr-8r^2 + (p+r)^{\nicefrac{-1}{2}}\frac{8r^2}{r^{\nicefrac{-1}{2}}} \right) 
\end{align}

\section{Application} \label{sec:app}



The solution $\bb{x}(\lambda)$ to the homotopy differential equation with initial value $\bb{x}(\lambda_0)$ can be constructed as
\begin{align}
    \bb{x}(\lambda) &= \bb{\Phi}(\lambda,\mu)\bb{x}(\mu) + \int_{\mu}^{\lambda}\bb{\Phi}(\lambda,\tau)\bb{b}(\tau)\diff{\tau} \\
    \bb{x}(\mu) &= \bb{\Phi}(\mu,\lambda_0)\bb{x}(\lambda_0) + \int_{\lambda_0}^{\mu}\bb{\Phi}(\mu,\tau)\bb{b}(\tau)\diff{\tau}
\end{align}
for some $\lambda_0\leq \mu \leq \lambda$. \hll{The interval $[0,1]$ can be split arbitrarily into $N$ parts with $\{\lambda_0=0 \leq \lambda_1 \leq \dots , \leq \lambda_N = 1\}$, and the above scheme applies with $N$ steps as:}
\begin{align} \label{eq:naedh_sol}
    \bb{x}(\lambda_i) = \bb{\Phi}(\lambda_i,\lambda_{i-1})\bb{x}(\lambda_{i-1}) + \int_{\lambda_{i-1}}^{\lambda_{i}}\bb{\Phi}(\lambda_i,\tau)\bb{b}(\tau)\diff{\tau} \, , (i = 1 \dots N) .
\end{align} 
This structure allows the handling of nonlinear systems, which is analogous to the numeric solution of the EDH filter.

To demonstrate the application of the analytic solution, the following filter implementations are tested in nonlinear estimation problems:
\begin{itemize}
    \item Extended Kalman filter (EKF)
    \item Exact Daum–Huang filter with Euler integration (EDH)
    \item Localized EDH with Euler integration (LEDH)
    \item EDH with analytic solution (A-EDH)
    \item EDH with N-step analytic solution (NA-EDH)
\end{itemize}
\hll{The general EDH-type filter structure is shown in Algorithm \ref{alg:edh}. The Bayesian particle filter update steps achieved by the EDH, LEDH, A-EDH, and NA-EDH are summarized in Algorithms \ref{alg:edh_euler}, \ref{alg:ledh_euler}, \ref{alg:a-edh}, and \ref{alg:na-edh}, respectively.}

\begin{algorithm}[htb]
\caption{Exact flow particle filter structure}
    \label{alg:edh}
    \begin{algorithmic}[1] 
        \State Initialize $\hat{\bb{x}}_0$ and $\bb{P}_{0}$ for the EKF and the particle set $\{\bb{x}^i_{0}\}_{i=1}^{N_p}$
        \For{$k = 1$ to $N_{k}$}
            \State \textbf{EKF prediction:}
            \State $(\hat{\bb{x}}_{k-1}$, $\bb{P}_{k-1}) \rightarrow (\hat{\bb{x}}_{k|k-1}$, $\bb{P}_{k|k-1})$ \Comment{Eq. \eqref{eq:EKFpredx}-\eqref{eq:EKFpredP}}
            \State \textbf{Particle prediction:}
            \State Draw from prior: $\bb{x}_{k|k-1}^{i} \sim \mathcal{N}(\bb{g}_k(\bb{x}_{k-1}^{i}), \bb{Q}_k)$ \Comment{Eq. \eqref{eq:particlePred}}
            \State \textbf{Particle flow update:} $\{\bb{x}^i_{k|k-1}\}_{i=1}^{N_p} \rightarrow \{\bb{x}^i_{k|k}\}_{i=1}^{N_p}$
            \State \hspace*{\algorithmicindent} $\bullet$ EDH with Euler integration \Comment Alg.~\ref{alg:edh_euler}
            \State \hspace*{\algorithmicindent} $\bullet$ Localized EDH with Euler integration \Comment Alg.~\ref{alg:ledh_euler}
            \State \hspace*{\algorithmicindent} $\bullet$ Analytic EDH \Comment Alg.~\ref{alg:a-edh}
            \State \hspace*{\algorithmicindent} $\bullet$ N-step analytic EDH\Comment Alg.~\ref{alg:na-edh}
            \State Compute point estimate: $\bar{\bb{x}}_{k} = \frac{1}{N_p} \sum_{i=1}^{N_p}\bb{x}_{k|k}^{i}$
             \State \textbf{EKF update:}
            \State $(\hat{\bb{x}}_{k|k-1}$, $\bb{P}_{k|k-1}) \rightarrow (\hat{\bb{x}}_{k|k}$, $\bb{P}_{k|k})$ \Comment{Eq. \eqref{eq:EKFres}-\eqref{eq:EKFupd}}
        \EndFor
    \end{algorithmic}
\end{algorithm}

\begin{algorithm}[htbp]
\caption{EDH update (Euler integration)}
    \label{alg:edh_euler}
    \begin{algorithmic}[1] 
        \Function{EDH}{$\bb{P}_{k|k-1},\{\bb{x}^i_{k|k-1}\}_{i=1}^{N_p}, z_k$} 
            \State Set $\bb{x}_0^{i}=\bb{x}^i_{k|k-1} \, , (i = 1\dots N_p)$
            \State Set $\lambda_0 = 0, \,  \Delta \lambda = 1/N_\lambda$
            \For{$j = 1$ to $N_{\lambda}$}
                \State $\lambda_j = \lambda_{j-1} + \Delta \lambda$
                \State Compute particle average: $\bar{\bb{x}}_{j} = \frac{1}{N_p}\sum_{i=1}^{N_p}\bb{x}_{j-1}^{i}$
                \State Linearize $h(\cdot)$ about $\bar{\bb{x}}_{j}$:  $\bb{H}(\bar{\bb{x}}_{j})$ \Comment{Eq. \eqref{eq:EKF_H}}
                \State Calculate $\bb{A}(\lambda)$ and $\bb{b}(\lambda)$ \Comment{Eq. \eqref{eq:A}-\eqref{eq:b}}
                    \For{$i = 1$ to $N_{p}$}
                        \State $\bb{x}_j^{i} = \bb{x}_{j-1}^{i} + (\bb{A(\lambda)} \bb{x}_{j-1}^{i} + \bb{b}(\lambda) ) \Delta \lambda$ \Comment{Eq. \eqref{eq:EDH_upd}}
                \EndFor
            \EndFor
            \State Updated particle set: $\bb{x}_{k|k}^{i} = \bb{x}^{i}_{\lambda=1} \, , (i = 1\dots N_p)$
         \State \textbf{return} $\{\bb{x}_{k|k}^{i}\}_{i=1}^{N_p}$
        \EndFunction
    \end{algorithmic}
\end{algorithm}

\begin{algorithm}[htbp]
\caption{LEDH update (Euler integration)}
    \label{alg:ledh_euler}
    \begin{algorithmic}[1] 
        \Function{LEDH}{$\bb{P}_{k|k-1},\{\bb{x}^i_{k|k-1}\}_{i=1}^{N_p}, z_k$} 
            \State Set $\bb{x}_0^{i}=\bb{x}^i_{k|k-1} \, , (i = 1\dots N_p)$
            \State Set $\lambda_0 = 0, \,  \Delta \lambda = 1/N_\lambda$
            \For{$j = 1$ to $N_{\lambda}$}
                \State $\lambda_j = \lambda_{j-1} + \Delta \lambda$
                \State Compute particle average: $\bar{\bb{x}}_{j} = \frac{1}{N_p}\sum_{i=1}^{N_p}\bb{x}_{j-1}^{i}$
                    \For{$i = 1$ to $N_{p}$}
                    \State Linearize $h(\cdot)$ about $\bb{x}_{j-1}^{i}$:  $\bb{H}(\bb{x}_{j-1}^{i})$ \Comment{Eq. \eqref{eq:EKF_H}}
                        \State Calculate $\bb{A}^{i}(\lambda)$ and $\bb{b}^{i}(\lambda)$ \Comment{Eq. \eqref{eq:A}-\eqref{eq:b}}
                        \State $\bb{x}_j^{i} = \bb{x}_{j-1}^{i} + (\bb{A}^{i}(\lambda) \bb{x}_{j-1}^{i} + \bb{b}^{i}(\lambda) ) \Delta \lambda$ \Comment{Eq. \eqref{eq:EDH_upd}}
                \EndFor
            \EndFor
            \State Updated particle set: $\bb{x}_{k|k}^{i} = \bb{x}^{i}_{\lambda=1} \, , (i = 1\dots N_p)$
         \State \textbf{return} $\{\bb{x}_{k|k}^{i}\}_{i=1}^{N_p}$
        \EndFunction
    \end{algorithmic}
\end{algorithm}

\begin{algorithm}[htbp]
\caption{Analytic EDH update}
    \label{alg:a-edh}
    \begin{algorithmic}[1] 
        \Function{A-EDH}{$\bb{P}_{k|k-1},\{\bb{x}^i_{k|k-1}\}_{i=1}^{N_p}, z_k$} 
            \State Set $\bb{x}_0^{i}=\bb{x}^i_{k|k-1} \, , (i = 1\dots N_p)$
                \State Compute particle average: $\bar{\bb{x}}_{j} = \frac{1}{N_p} \sum_{i=1}^{N_p}\bb{x}_{j-1}^{i}$
                \State Linearize $h(\cdot)$ about $\bar{\bb{x}}_{j}$:  $\bb{H}(\bar{\bb{x}}_{j})$ \Comment{Eq. \eqref{eq:EKF_H}}
                \State Calculate $\bb{A}(\lambda)$ and $\bb{b}(\lambda)$ \Comment{Eq. \eqref{eq:A}-\eqref{eq:b}}
                    \For{$i = 1$ to $N_{p}$}
                        \State Calculate $\bb{x}^{i}_{\lambda=1}$ in one step: \Comment{Eq. \eqref{eq:aedh_sol}}  
                        \State $\bb{x}^{i}_{\lambda=1} = \bb{\Phi}(1,0)\bb{x}_0^{i} +\sum_{l=0}^{4} \bb{\Psi}_l(1,0)$ \Comment{Eq. \eqref{eq:sol_homFI}-\eqref{eq:sol_homb4}}
                \EndFor
            \State Updated particle set: $\bb{x}_{k|k}^{i} = \bb{x}^{i}_{\lambda=1} \, , (i = 1\dots N_p)$
         \State \textbf{return} $\{\bb{x}_{k|k}^{i}\}_{i=1}^{N_p}$
        \EndFunction
    \end{algorithmic}
\end{algorithm}

\begin{algorithm}[htbp]
\caption{N-step analytic EDH update}
    \label{alg:na-edh}
    \begin{algorithmic}[1] 
        \Function{NA-EDH}{$\bb{P}_{k|k-1},\{\bb{x}^i_{k|k-1}\}_{i=1}^{N_p}, z_k$} 
            \State Set $\bb{x}_0^{i}=\bb{x}^i_{k|k-1} \, , (i = 1\dots N_p)$
            \State Set $\lambda_0 = 0, \,  \Delta \lambda = 1/N_\lambda$
            \For{$j = 1$ to $N_{\lambda}$} \Comment{$N$-step scheme from Eq. \eqref{eq:naedh_sol}}
                \State $\lambda_j = \lambda_{j-1} + \Delta \lambda$
                \State Compute particle average: $\bar{\bb{x}}_{j} = \frac{1}{N_p} \sum_{i=1}^{N_p}\bb{x}_{j-1}^{i}$
                \State Linearize $h(\cdot)$ about $\bar{\bb{x}}_{j}$:  $\bb{H}(\bar{\bb{x}}_{j})$ \Comment{Eq. \eqref{eq:EKF_H}}
                    \For{$i = 1$ to $N_{p}$} 
                       \State Calculate $\bb{x}^{i}_{j}$ substeps: \Comment{Eq. \eqref{eq:aedh_sol}}
                       \State $\tilde{\bb{x}}^{i}_{j} = \bb{\Phi}(\lambda_j,\lambda_{j-1})\bb{x}_{j-1}^{i}$ \Comment{Eq. \eqref{eq:Phi}}
                       \State $\bb{x}^{i}_{j} = \tilde{\bb{x}}^{i}_{j} + \sum_{l=0}^{4} \bb{\Psi}_l(\lambda_j,\lambda_{j-1})$ \Comment{Eq. \eqref{eq:b0}, \eqref{eq:b1}, \eqref{eq:b2}, \eqref{eq:b3}, \eqref{eq:b4}}
                \EndFor
            \EndFor
            \State Updated particle set: $\bb{x}_{k|k}^{i} = \bb{x}_{N_\lambda}^{i} \, , (i = 1\dots N_p)$
         \State \textbf{return} $\{\bb{x}_{k|k}^{i}\}_{i=1}^{N_p}$
        \EndFunction
    \end{algorithmic}
\end{algorithm}

The first nonlinear model that is commonly used for filter evaluation \cite{arulampalam2002tutorial}, is one-dimensional:
\begin{align} \label{eq:1Dx}
    x_k &= \frac{x_{k-1}}{2} + \frac{25x_{k-1}}{1+x_{k-1}^2}+8\cos{1.2k} + w_k \\
    z_k &= \frac{x_k^2}{20} + v_k \, , \label{eq:1Dz}
\end{align}
where the variance for the AWGN $w_k$ is $Q=10$ and for $v_k$ is $R=0.1$.

The second nonlinear model has a multidimensional state vector $\bb{x}$, a linear, fully coupled, stable motion model, and a nonlinear measurement model in the form
\begin{align}
    \bb{x}_{k+1} &= \bb{F} \bb{x}_{k} + \bb{w}_k \\
    z_k &= \bb{x}_{k}^\top \bb{x}_{k} + v_k
\end{align}
To run numerous Monte Carlo (MC) simulations, the system parameters are generated randomly. $\bb{F}$ is generated as $\bb{F} = \bb{T}_F\bb{U}\bb{T}_F^{-1}$ where $\bb{U}$ is a diagonal matrix with random negative eigenvalues. The process noise covariance is generated as $\bb{Q} =\bb{T}_Q\bb{T}_Q^{\top}$. The matrices $\bb{T}_F$ and $\bb{T}_Q$ have random positive entries. The used random numbers are uniformly distributed in the open interval (0,1). The measurement noise variance is $R=5$.

To evaluate filter performance, 100 MC runs were performed for every filter configuration. For the particle flow filters, the used particle numbers are 10, 50, 100, and 500. For the EDH and NA-EDH filters, 10 steps were used. The simulations took 100 steps with stepsize 1. For the multidimensional model, 10, 50, and 100 dimensions were used. The simulation was implemented in MATLAB.

\begin{figure}
    \centering
    \includegraphics[width=0.85\linewidth]{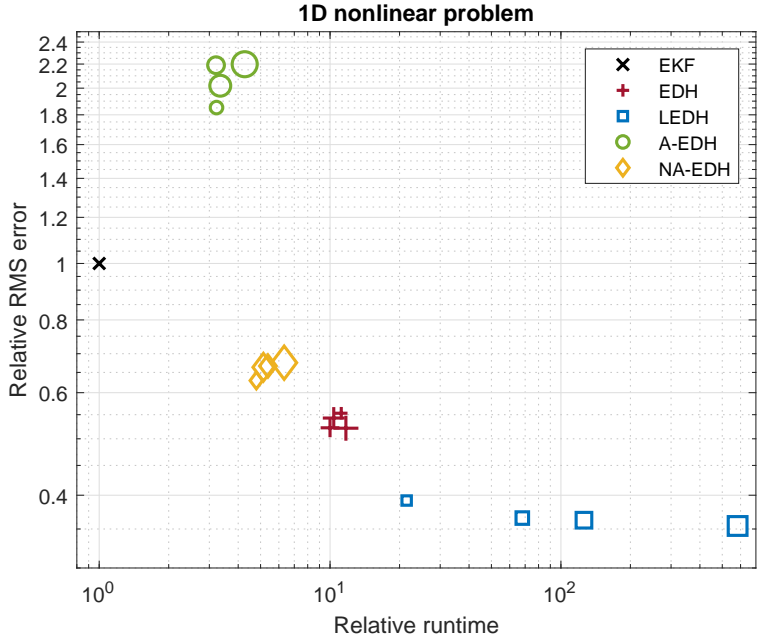}
    \caption{Results for the one dimensional model}
    \label{fig:1D}
\end{figure}

\begin{figure}
    \centering
    \includegraphics[width=0.85\linewidth]{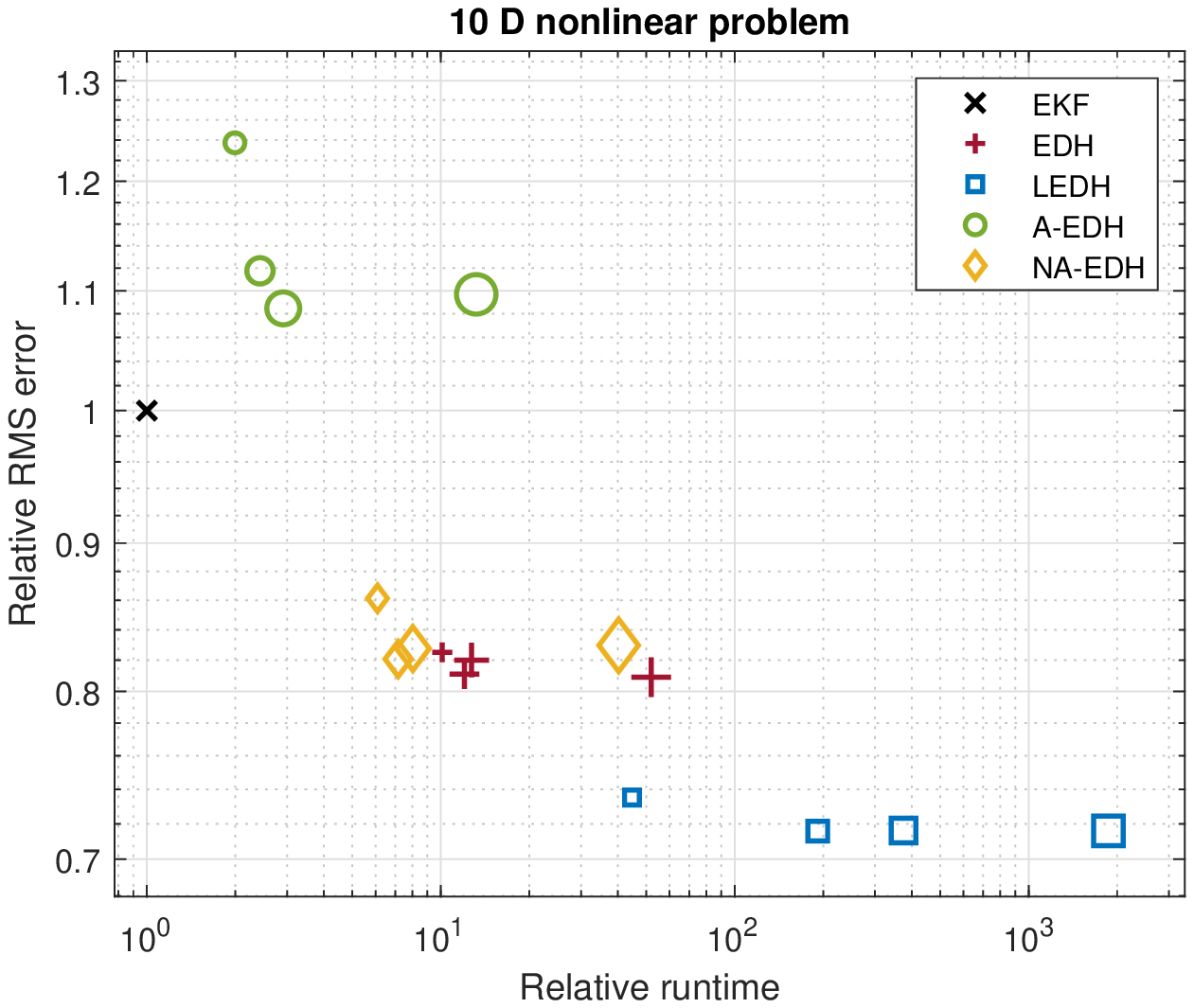}
    \caption{Results for the 10 dimensional model}
    \label{fig:10D}
\end{figure}

\begin{figure}
    \centering
    \includegraphics[width=0.85\linewidth]{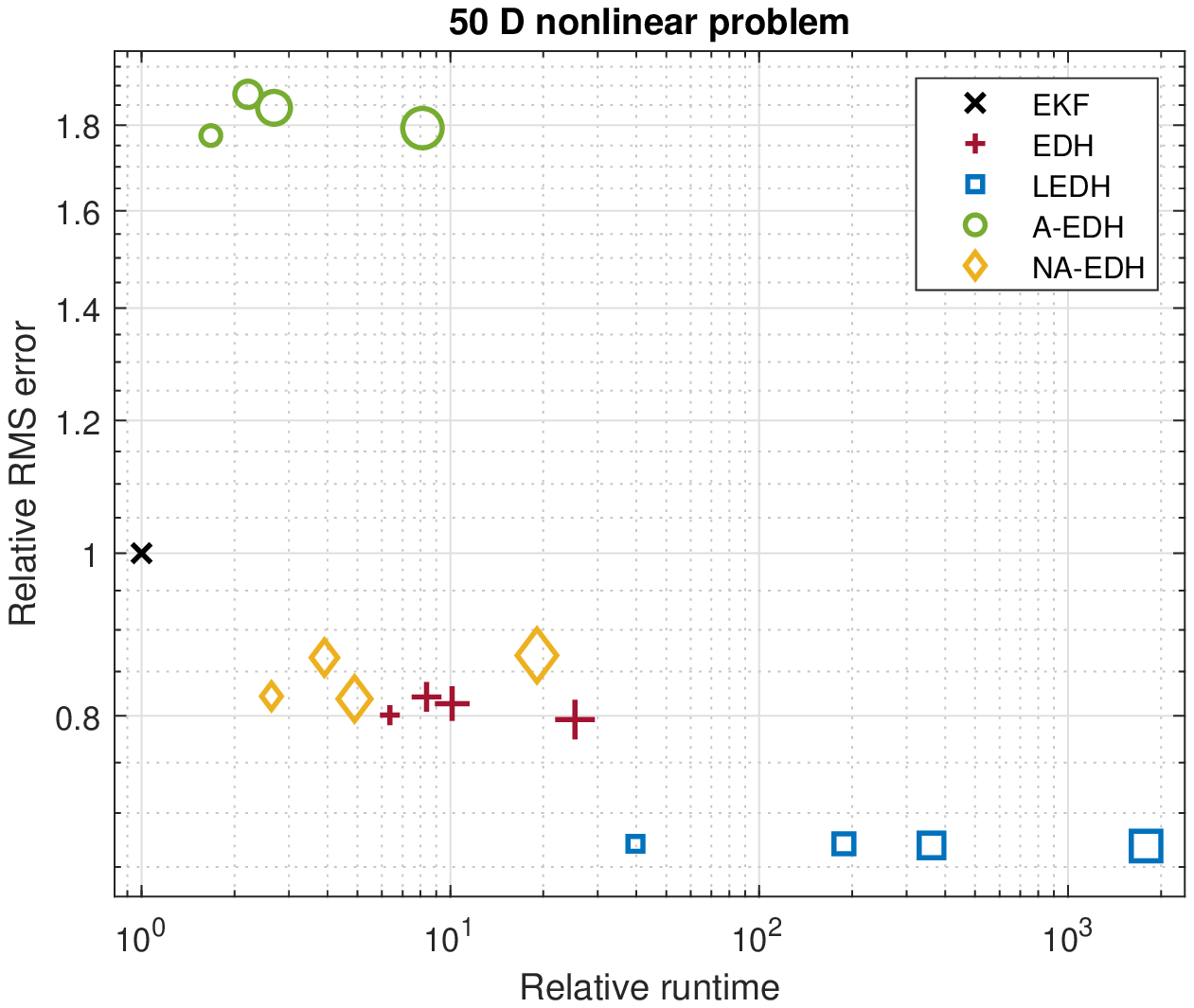}
    \caption{Results for the 50 dimensional model}
    \label{fig:50D}
\end{figure}

\begin{figure}
    \centering
    \includegraphics[width=0.85\linewidth]{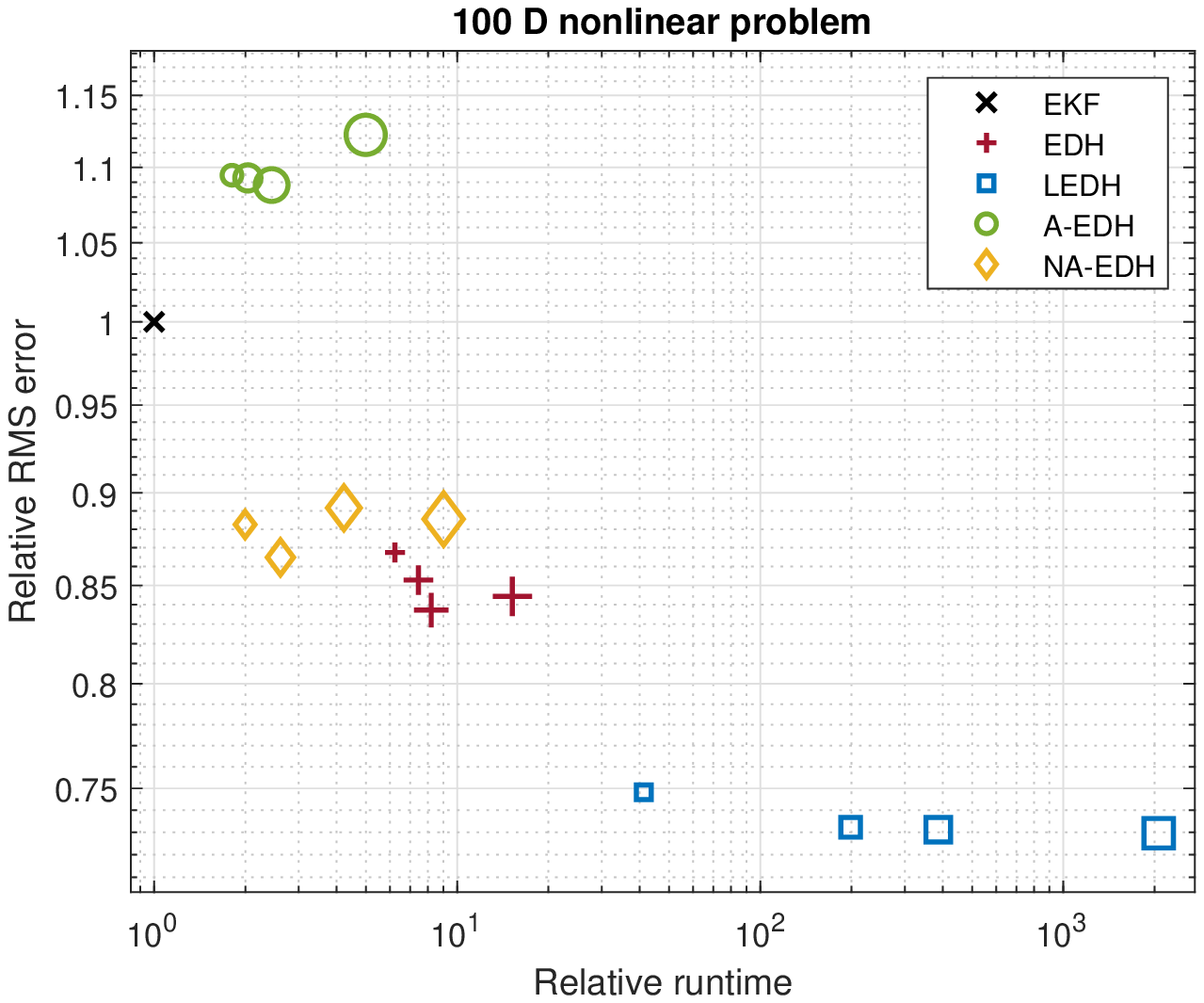}
    \caption{Results for the 100 dimensional model}
    \label{fig:100D}
\end{figure}

The results are evaluated relative to the EKF performance. The root mean squared error (RMSE) values of the estimations are averaged over the 100 MC runs and normalized by the EKF RMSE value. The average filter runtimes are also normalized by the EKF runtime. These relative values plotted against each other help to visualize the filters' performance. Fig. \ref{fig:1D} shows the result for the one dimensional model while Fig. \ref{fig:10D}, \ref{fig:50D}, and \ref{fig:100D} are for the multidimensional cases. Larger plotmarkers indicate more particles. For reference, the absolute performance values for the 100 dimensional case are shown in Table \ref{tab:perf}.

It clearly stands out that using the analytic solution in one step is not favorable. The reason is that the solution is only valid for a linear ODE, and using it in this form could be approximated by \eqref{eq:EDH_upd} with a drift not depending on the position. Instead of $\bb{f}(\overline{\bb{x}}_{n-1}^{i}, \lambda)$ one needs $\bb{f}(\overline{\bb{x}}_{k|k-1}, \lambda)$ thus
\begin{equation} 
    \bb{x}_n^{i} = \bb{x}_{n-1} + \bb{f}(\overline{\bb{x}}_{k|k-1}, \lambda) \Delta \lambda \, \, \, \, \,  (n = 1 \dots N) \, ,
\end{equation}
is the Euler approximation for the analytic solution. It can be said that with the analytic solution we sacrifice the spatial dependence of the flow to gain resolution in $\lambda$. This approach is not fruitful, and the multi-step solution can be used instead, which inherits the spatial dependence of the flow from \eqref{eq:EDH_upd}. As one would expect, the performance of the NA-EDH is comparable to the EDH with numeric integration and needs less computation.

The completely localized filter provides the best performance for a huge computational cost. It is reasonable to say that some amount of localization is certainly needed to get a satisfactory performance.

\begin{table}[htbp]
\caption{RMSE and Runtime (ms) values for the 100 dimensional model using 10, 50, 100, and 500 particles}
\label{tab:perf}
\centering
\setlength\extrarowheight{2.5pt}
\begin{tabular}{@{}lccccc@{}}
\toprule
Performance & EKF & EDH & LEDH & A-EDH & NA-EDH \\
\midrule
$\text{RMSE}_{10}$ & 542 & 470 & 405 & 593 & 478 \\
$\text{TIME}_{10}$ & 34.77 & 212.8 & 1408.8 & 61.66 & 68.14 \\
$\text{RMSE}_{50}$ & 542 & 462 & 397 & 592 & 469 \\
$\text{TIME}_{50}$ & 34.07 & 254.1 & 6781.6 & 69.58 & 89.12 \\
$\text{RMSE}_{100}$ & 542 & 454 & 396 & 589 & 483 \\
$\text{TIME}_{100}$ & 33.36 & 280.3 & 13194 & 83.37 & 144.3 \\
$\text{RMSE}_{500}$ & 542 & 457 & 395 & 608 & 480 \\
$\text{TIME}_{500}$ & 34.36 & 519.3 & 70401 & 170.33 & 307.9 \\
\bottomrule
\end{tabular}
\end{table}


\section{Conclusion} \label{sec:conc}
The restriction for scalar measurement may seem too limiting; however, there are cases, even in transportation, when a single scalar value, e.g., the road slope, can help estimate important quantities \cite{nemeth2022impact}.

Several directions for further development of the analytic solution based EDH filter can be appointed. First, the equations in the solution are not optimized for computation. Additional simplifications may lighten the computational needs. Second, the amount of localization in the drift term needs a balance with the number of particles and steps in $\lambda$.
Lastly, the most obvious direction is to generalize the solution to the vector measurement case.

\hl{Alternatively, one might attempt to process a vector measurement sequentially as scalars. Processing radar measurements sequentially in a preferred order and also reducing the linearization error of the EKF state update to third-order is discussed in \cite{miller1982nonlinear}. A newer method called the extended preferred ordering theorem completely abolishes the preferred ordering for sequential measurement processing, as reported in \cite{leskiw2009extended}.}

\hl{The standard formulation of the exact flow particle filter discussed in this work assumes Gaussian distributions. In the case of non-Gaussian distributions, approximations can be used. One example is the approach of Kamen for designing an extended Kalman filter with symmetric measurement equations for a multi-target estimation problem \cite{kamen1989multiple}. Derivations for nonlinear transformations of Gaussians for these type of problems can be found in \cite[pp.~298-301]{kamen1999introduction}, in the papers of Leven \cite{leven2009unscented,leven2005multiple,leven2006thesis} or in \cite{baum2013kernel}.
There are algorithms designed for cases when the noise terms, and perhaps the initial state also, are of Gaussian sum type. The well-established Gaussian sum approach of Sorenson and Alspach \cite{sorenson1971recursive} has been used to implement the exact flow particle filter in \cite{pal2017gaussian,pal2018particle,khan2016log}. It may be worthwhile to investigate whether other similar approaches, such as those presented in \cite{masreliez1975approximate} and \cite{wu1999suboptimal}, could be applied effectively with a particle flow filter.}

\hll{The particle flow implementation of the Bayesian update step does not use weights or a resampling strategy. Particle degeneracy and collapse are also not an issue thus a regularization step is not part of a particle flow filter. Nonetheless, the optimal choice of step sizes in $\lambda$ is of great interest \cite{mori2016adaptive}. Besides the equal or exponentially increasing step sizes that apply for all particles, unique steps can be assigned to every particle based on their velocities \cite{mori2016adaptive, crouse2019particle}. Further inspiration can be gained to optimize step sizes from the more refined methods presented in \cite{musso2001improving}.}

\hl{Currently the stochastic particle flow based on Gromov's method, introduced in {\cite{daum2016gromov}}, gives the best accuracy  {\cite{pal2019particle,daum2018newtheory, crouse2019particle}}. The implementation is, however, challenging for a general distribution represented by particles as a Dirac-sum, thus certain tricks are needed \cite{daum2009gradient}, \cite[p. 106]{evensen2022data}.}
In \cite{dai2021new} a unified discussion of particle flows parameterized by a homogeneous diffusion matrix is presented. On the question of modeling inhomogeneous diffusion, van Kampen has drawn the conclusion that "no universal form of the diffusion equation exists, but each system has to be studied individually" \cite{van1987diffusion}. This is satisfactory since the diffusion equation is phenomenological that tries to capture the complex interactions of molecules, which would be described in detail by physical kinetics. Contrarily, for the particle flow, there are no underlying physical laws, and we are free to select or design diffusion equations that best serve our needs.


\section{Acknowledgement}

The research was supported by the European Union within the framework of the National Laboratory for Autonomous Systems. (RRF-2.3.1-21-2022-00002) \\The research reported in this paper is part of project no. BME-NVA-02, implemented with the support provided by the Ministry of Innovation and Technology of Hungary from the National Research, Development and Innovation Fund, financed under the TKP2021 funding scheme.








\end{document}